\documentclass[acmsmall,screen,nonacm]{acmart} 
\usepackage{url}
\usepackage{verbatim}
\usepackage{graphicx}
\usepackage{color}

\setcopyright{none}
\renewcommand\footnotetextcopyrightpermission[1]{} 
\AtBeginDocument{%
  }

\begin{document}

\title{Prospects for inconsistency detection using large language models and sheaves}

%
\author{Steve Huntsman}
\email{sch213@nyu.edu}
\orcid{0000-0002-9168-2216}
%
\author{Michael Robinson}
\email{michaelr@american.edu}
\orcid{0000-0003-0766-3301}
\author{Ludmilla Huntsman}
\email{ludmilla@cogsecalliance.org}
\orcid{0009-0002-6599-0941}


\begin{abstract}
We demonstrate that large language models can produce reasonable numerical ratings of the logical consistency of claims. We also outline a mathematical approach based on sheaf theory for lifting such ratings to hypertexts such as laws, jurisprudence, and social media and evaluating their consistency globally. This approach is a promising avenue to increasing consistency in and of government, as well as to combating mis- and disinformation and related ills.
\end{abstract}


\maketitle

\section{\label{sec:introduction}Introduction}

\begin{figure}[h]
  \centering
  \fbox{\includegraphics[trim = 83mm 100mm 75mm 115mm, clip, width=.167\columnwidth,keepaspectratio]{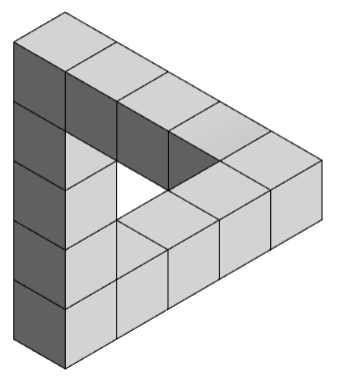}}
  \fbox{\includegraphics[trim = 83mm 100mm 75mm 115mm, clip, width=.167\columnwidth,keepaspectratio]{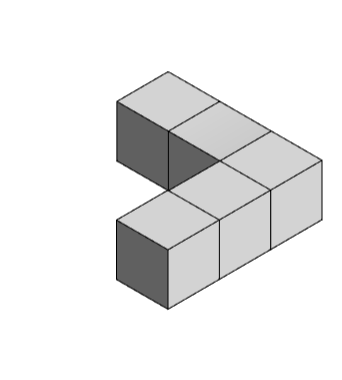}}
  \fbox{\includegraphics[trim = 83mm 100mm 75mm 115mm, clip, width=.167\columnwidth,keepaspectratio]{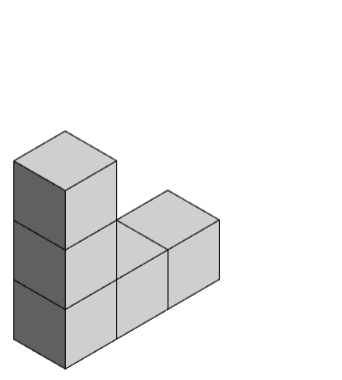}}
  \fbox{\includegraphics[trim = 83mm 100mm 75mm 115mm, clip, width=.167\columnwidth,keepaspectratio]{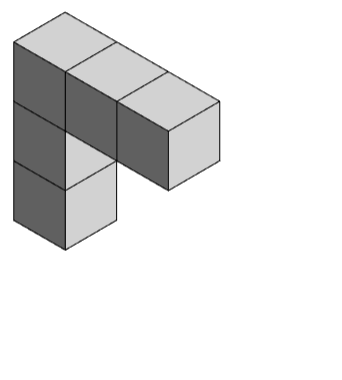}}
  \fbox{\includegraphics[trim = 83mm 100mm 75mm 115mm, clip, width=.167\columnwidth,keepaspectratio]{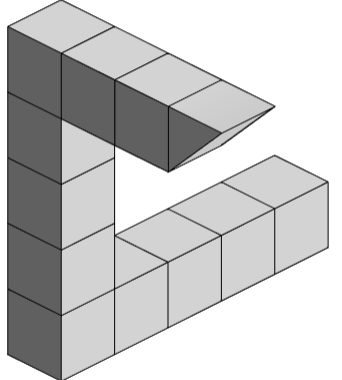}}
  \caption{Sheaf cohomology is used in \cite{penrose1992cohomology} (and its reproduction as Example 231 of \cite{rosiak2022sheaf}) to detail how the Penrose triangle in the left panel cannot be realized by consistently gluing together local data suggested by the middle three panels, i.e., the cubes at the ends of the three L shapes. The calculation also clarifies how ambiguity of perspective in a two-dimensional drawing is essential to the illusion of global consistency if the shape's connectivity is actually taken to be that of a triangle instead of as shown in the right panel.
  }
  \Description{Sheaf cohomology describes obstructions to global consistency of data.}
  \label{fig:triangle}
\end{figure}

Imagine that administrators, legislators, and judges could be respectively nudged \cite{thaler2021nudge} towards producing more consistent bodies of policy, law, and jurisprudence \cite{cyrul2013consistency,kahneman2021noise}; and that common mis- and disinformation \cite{chen2023combating,cipers2023government,jiang2023disinformation,wef2024global}, doublethink \cite{petrovic2022thinking}, hypocrisy \cite{galitsky2021truth}, and bullshit \cite{frankfurt2005bullshit,frankfurt2010truth,bergstrom2023chatgpt,littrell2023bullshit} in public life could be reliably automatically detected and forensically detailed in real time. This Utopia---in which Brandolini's law that ``the amount of energy needed to refute bullshit is an order of magnitude bigger than that needed to produce it'' \cite{bergstrom2021calling} would no longer hold---may not be as fanciful as it seems. Large language models (LLMs) appear to have the ability to solve a key problem of making quantitative local judgments about human language in hypertexts whose global consistency can then be evaluated using mathematical techniques from \emph{sheaf theory} \cite{ghrist2014elementary,rosiak2022sheaf}, as schematically indicated in Figure \ref{fig:triangle}.

Briefly put, sheaves describe how to consistently restrict global data in order to obtain local data. For example, imagine a collection of identical pictures that are cut into pieces differently. Each of the resulting fragments is local data restricted from the global data of the original picture.  This may sound trivial, but the machinery works both ways: given a collection of picture fragments, we can arrange and glue them together--\emph{including in different ways than before, provided they overlap consistently}--to produce copies of the original picture again. That is, sheaves equivalently describe how to glue local data together to form global data.

Although the idea of using sheaves to represent inconsistent information in texts is more than five years old \cite{zadrozny2018sheaf}, this (to our knowledge only) prior approach in the literature is based on an algebraic---and therefore somewhat brittle---representation of parametric natural language sentences (e.g., possibly conflicting bits of quantitative medical advice). A topological representation of the sort sketched below (which differs at a fundamental level from a topological \emph{analysis} of a vector representation such as \cite{wu2022topological}) would have the advantage of tolerating benign errors in a graceful way. The advent of LLMs suggests an elementary device for completely avoiding nontrivial mathematics at the level of individual claims in text\footnote{In our experience, ChatGPT is quite good at extracting the most important claims in a text, but we have not rigorously evaluated its performance here, and a tool such as \cite{gupta2021lesa} can be considered for this task as well.}
and instead relying on a separation of concerns in which inconsistency is represented locally using LLM outputs and globally using the machinery of sheaves. Moreover, recent advances in tooling for sheaves suggests that they gracefully tolerate irrelevant differences in their inputs \cite{robinson2019hunting}, so the well-known variability in LLM outputs may not pose a major concern.

Because of the ambition of our goals, it is important to be clear about limitations of technology as well as prospects. For example, the training of LLMs entails bounds on their ability to deal with text pertaining to sufficiently esoteric, rare, or new information that does not have adequate representation and context in their training corpora.\footnote{W. Zadrozny has pointed out [personal communication] that biases, false background knowledge, etc. might prevent an LLM from reasoning correctly. While on one hand this is a ``turtles all the way down'' problem, using an approach such as we propose in the course of LLM training also suggests a possible solution.} As another example, it is entirely reasonable to want any system that judges the consistency of claims to work in an online setting, i.e., to be able to incrementally accept new claims. However, the order in which these new claims arrive may have a substantial impact, and we are not aware of a principled way to handle temporal knowledge bases that is fully self-consistent. Because LLM outputs are probabilistically sampled and hence intrinsically variable, it is not unreasonable to suspect that any system that relies upon them for consistency measurement may exhibit instability over time. Claims will need to be carefully time-ordered and curated from the start. \emph{Who claims what and when they claim it} is critical for assessing consistency in a network of claims.  

Analogous challenges also apply to implementing a system that might achieve our goals. The sheer volume of manifestly inconsistent information, and the rates at which it is produced and consumed, pose an extremely serious challenge to honest, effective government and to healthy public life \cite{frankfurt2010truth}, but also to system scaling. Besides fundamental issues with scaling in volume (hence also cost) and time, it is also necessary to simultaneously achieve reliability from LLM outputs. While as we show in \S \ref{sec:local} the latter can be achieved by repetition and/or variation of inputs and averaging of outputs (cf. \cite{wang2022self}), this imposes additional scaling demands and also demands careful statistical analysis. Some of these scaling demands can be ameliorated using lightweight and/or fine-tuned models and triaging inadequate (e.g., bimodal) preliminary results for handling by more powerful models, as well as by leveraging parallelism, but new insights will still be necessary for any practical deployment. Even before any of that, many algorithmic and engineering challenges common to any new technological development must be overcome. Nevertheless, we believe there is a reasonably clear path towards a proof of concept implementation, in part because recent advances in computational topology have yielded practical implementations of efficient algorithms, and of course also in large part due to the advent of LLMs.

Before detailing our results about local consistency of claims in \S \ref{sec:local}, our proposed framework for determining global consistency of claims in \S \ref{sec:global}, and remarks on demonstration and deployment in \S \ref{sec:deployment}, we first make a few general remarks. Although it hardly makes sense to try to provide detailed or truly authoritative references for LLMs given the rate at which the technology is evolving, the technological substrate of transformers is discussed in \cite{vaswani2017attention,arora2020theory,turner2023introduction}; the details of domain-specific and multimodal LLMs are respectively discussed in \cite{taylor2022galactica} and \cite{team2023gemini}; and emergent abilities of LLMs are discussed in \cite{wei2022emergent}, with a plausible theory for emergence provided in \cite{arora2023theory}. Having made this nod towards the actual details of LLM design, operation, and behavior, we may engage in some obvious anthropomorphism throughout for the sake of readability. We note finally in reference to possible applications of our ideas in the legal realm that LLMs have been applied to produce legal reasoning \cite{deng2023syllogistic,jiang2023legal,kang2023can,yu2023exploring}, and that \texttt{GPT-4} passed the Uniform Bar Examination \cite{katz2023gpt}.

\section{\label{sec:local}Judging local consistency of claims}

\begin{figure}[h]
  \centering
  \includegraphics[trim = 11.5mm 50mm 23.5mm 60mm, clip, width=.8\columnwidth,keepaspectratio]{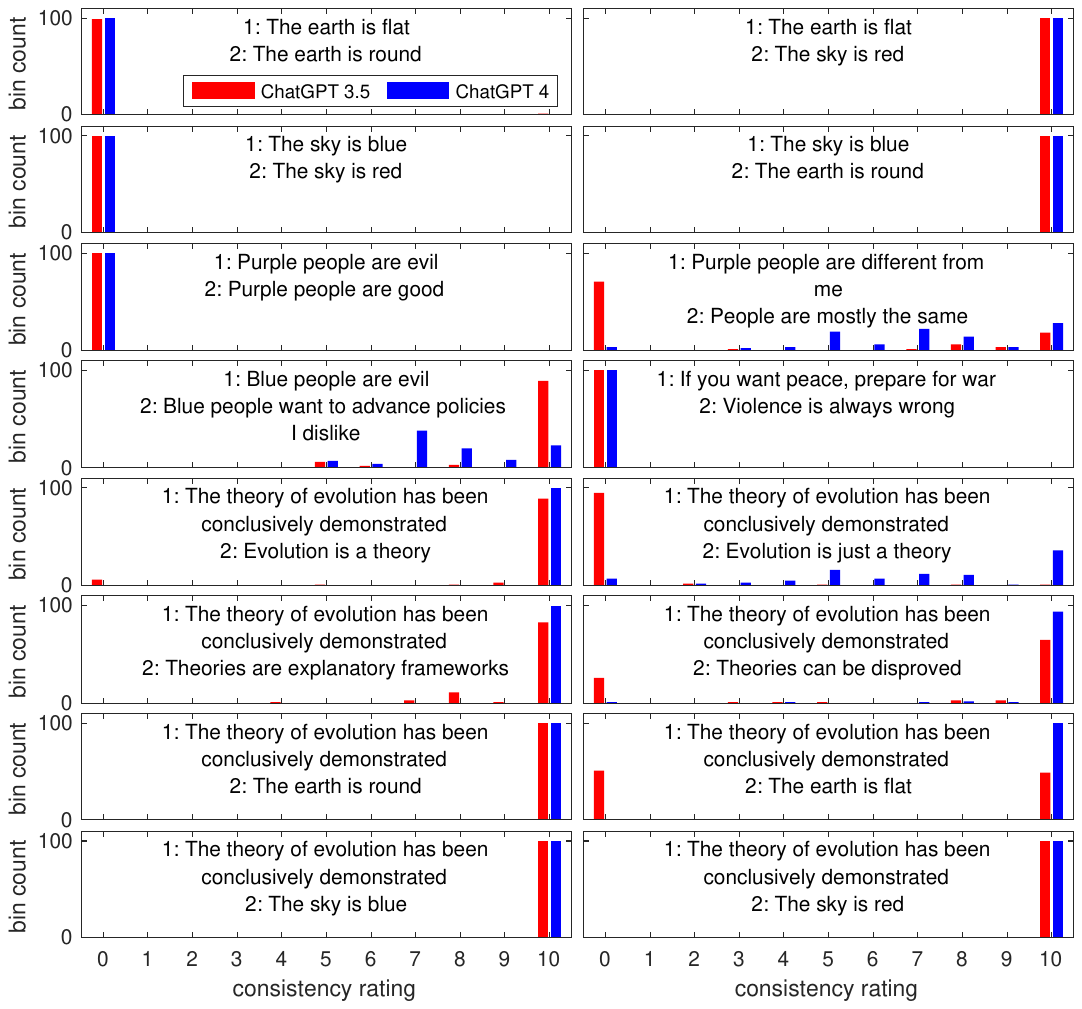}
  \caption{\texttt{ChatGPT} can reliably quantify the logical consistency of claims, extending the binary inconsistency detection demonstrated in \cite{mundler2023self} and \cite{li2023contradoc}. Here, we show histograms of numerical consistency ratings produced by two versions of \texttt{ChatGPT} with default configurations in response to the initial prompt in \S \ref{sec:prompt} followed by the two claims indicated in each figure panel. For each pair of claims, we gave \texttt{ChatGPT} 3.5 and 4 the same prompt $N = 100$ times and extracted the numerical consistency rating it produced at the end of an explanation of the logical consistency of these claims (in just a single case, \texttt{ChatGPT} 3.5 failed to produce a rating at the end of its reply; \texttt{ChatGPT} 4 never failed to). NB. The prompt in \S \ref{sec:prompt} includes the claim pairs $(\texttt{The earth is flat},\texttt{The sky is red})$ and $(\texttt{Purple people are evil},\texttt{Purple people are good})$ as its \emph{only} two examples for few-shot learning.
  }
  \Description{ChatGPT can reliably quantify the logical consistency of claims.}
  \label{fig:local}
\end{figure}

Because LLMs have strong logical reasoning abilities \cite{liu2023evaluating}, they are able to efficiently detect manifest logical inconsistencies in text \cite{mundler2023self,li2023contradoc}.
\footnote{See also \url{https://chatprotect.ai/}.} 
\footnote{LLMs also generally prefer factually consistent continuations of their inputs \cite{tam2023evaluating}.} 
Here, we show that LLMs can go beyond this by providing reasonable fine-grained quantifications of logical inconsistency of short texts such as claims. Figure \ref{fig:local} shows the results of an experiment in which we asked \texttt{ChatGPT} versions 3.5 (specifically, the \texttt{turbo} variant) and 4 to explain the logical relationship between various pairs of claims and subsequently numerically rate the logical consistency of those claims on a scale from 0 (inconsistent) to 10 (consistent). 
\footnote{We actually ran this experiment twice, \emph{obtaining virtually indistinguishable distributions}. In the first run, minor errors (found late in writing) in saving and processing the data precluded us from treating that data as authoritative. The scripts used and data produced in the second run are available at \url{https://github.com/SteveHuntsman/ProspectsForInconsistencyDetection}.}



Although the prompt used for this purpose (reproduced in \S \ref{sec:prompt}) required considerable experimentation, the results appear encouraging. Even the bimodal responses of \texttt{ChatGPT} 3.5 arise from what could be termed defensible interpretations were a human to produce them. For example, its first consistency rating of 0 for the claim pair
$$(\texttt{The theory of evolution has been conclusively demonstrated},\texttt{The earth is flat})$$
was extracted from the following response that spot checking suggests is broadly typical for that rating:

\begin{quote}
\texttt{If the theory of evolution has been conclusively demonstrated, it implies that the theory is supported by a significant body of evidence and scientific consensus. On the other hand, the claim that the earth is flat is contrary to overwhelming scientific evidence, which demonstrates that the earth is a spheroid. Therefore, if the theory of evolution has been conclusively demonstrated, it is logical to conclude that the earth is not flat.\textbackslash n\textbackslash nThe logical relationship between the two claims is contradictory. If the first claim is true, then the second claim is false, and if the second claim is true, then the first claim is false. Thus, the consistency rating of these claims is 0.}
\end{quote}
That is, the model appears to be:
\begin{itemize}
	\item[i)] identifying that each claim relates to scientific knowledge; 
	\item[ii)] identifying that one claim is consistent with scientific knowledge, and that the other claim is inconsistent with scientific knowledge; 
	\item[iii)] combining the latter identification with a textual representation of a basic logical argument to arrive at a logically misguided but still defensible consistency rating of the claims.
\end{itemize}
Meanwhile, \texttt{ChatGPT} 4 makes no such errors of logic that are evident in ratings or spot-checked explanations, presumably due to a more nuanced internal representation of the prompt. \texttt{ChatGPT} 4 also appears to produce more nuanced evaluations of logical consistency of claims, as evidenced by the increased variance in rating with more ambiguously related claim pairs. In any event, bimodality or even large variance from a LLM can serve as an indicator to perturb a prompt and/or hand off to a more powerful model until this is either resolved or imputed to fundamental ambiguity.

\subsection{\label{sec:toyExample}A harder example\protect\footnote{We thank James Fairbanks for suggesting this example; cf. the second, third, and fourth panels of Figure \ref{fig:local}.}}

\begin{figure}[h]
  \centering
  \includegraphics[trim = 11.5mm 115mm 23.5mm 125mm, clip, width=.8\columnwidth,keepaspectratio]{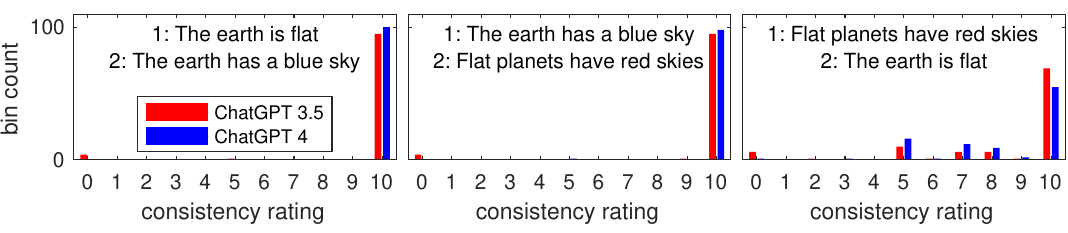}
  \caption{As in Figure \ref{fig:local}, but for the claim pairs of \S \ref{sec:toyExample}.
  }
  \Description{ChatGPT can resolve simple paradoxes.}
  \label{fig:example}
\end{figure}

With Figure \ref{fig:triangle} in mind, we also performed a variant of the experiment described in Figure \ref{fig:local} but with the following three pairs of claims:
$$(\texttt{The earth is flat},\texttt{The earth has a blue sky});$$
$$(\texttt{The earth has a blue sky},\texttt{Flat planets have red skies});$$
$$(\texttt{Flat planets have red skies},\texttt{The earth is flat}).$$
If all three pairs of claims were perfectly consistent, this would entail that blue and red are the same. Figure \ref{fig:example} shows how this inconsistency can be detected.

When \texttt{ChatGPT} assigns lower consistency ratings to the third claim pair, this is because it correctly identifies a conditional relationship between the claims.\footnote{See data at \url{https://github.com/SteveHuntsman/ProspectsForInconsistencyDetection}.}
\footnote{The implicit universal (i.e., ``for all'') quantifier over planets may be a particular sticking point. LLMs have been observed to have difficulty with quantifiers \cite{gupta2023probing}.}
As its responses detail, \texttt{ChatGPT} 3.5 identifies this relationship in instances that it rates below 10, though it makes logical missteps at the same time. Meanwhile, most of the time the explanations and ratings from \texttt{ChatGPT} 4 are reasonable. For example, its first explanation with a rating of 5 is
\begin{quote}
\texttt{If we assume the first claim is true that flat planets have red skies, it does not provide any explicit information about whether the second claim - that the Earth is flat - is true or false. However, if we were to assume the second claim is true while holding the first claim true, it might imply that the Earth should have a red sky, which is not the subject of either claim. These claims are neither completely consistent nor completely inconsistent – they form a possible but not certain logical chain. As such, I would give a consistency rating of 5.}
\end{quote}
The implicit inconsistency is thus detectable and traceable to at least one of the claims in the third pair. Still, claims involving quantifiers are comparatively tricky.\footnote{It is possible to lay an insidious trap using the Brandenburger-Keisler paradox \cite{brandenburger2006impossibility,pacuit2007understanding}. We thank Daniel Rosiak for pointing this out. \texttt{ChatGPT} generally rates the consistency of the claims \texttt{Alice believes that Bob assumes that Alice believes that Bob’s assumption is wrong} and \texttt{Alice believes that Bob’s assumption is wrong} highly. Replacing the latter claim with \texttt{Alice does not believe that Bob’s assumption is wrong} has the reverse effect: the resulting consistency ratings are usually zero. In any event, neither version of \texttt{ChatGPT} appears capable of recognizing the paradox, and the reasoning \texttt{ChatGPT} provides in its ratings is unsurprisingly always flawed. Still, we must point out that this paradox was discovered fairly recently and has an intricate resolution, viz., that some descriptions of beliefs are impossible to represent. It seems unreasonable to expect an LLM to compete with a professional logician in evaluating the consistency of claims. Future AI systems may not be at such a disadvantage.
}

\section{\label{sec:global}Judging global consistency of claims}

Even if we take \S \ref{sec:local} as adequate preliminary evidence that LLMs can numerically rate the logical consistency of pairs or small tuples 
of claims, how can we lift such local data to assess globally interlinked claims in government policy, law, and jurisprudence; and for that matter also in (mostly social) media? The key, anticipated by \cite{zadrozny2018sheaf} (but in that work limited to text that admits parametrized algebraic representations by virtue of its quantitative content) and previewed in \S \ref{sec:introduction}, is the notion of \emph{sheaf} that precisely describes the mathematics of ``gluing'' local data or \emph{sections} together \cite{ghrist2014elementary,hansen2021gentle,rosiak2022sheaf}. This can be effected computationally using a framework such as \texttt{PySheaf}\footnote{\url{https://github.com/kb1dds/pysheaf}}
or \texttt{StructuredDecompositions.jl}\footnote{\url{https://github.com/AlgebraicJulia/StructuredDecompositions.jl}}
\cite{althaus2023compositional}.

The archetypal example is the set of continuous functions on the real line: the restriction of a continuous function to the intersection of two open intervals agrees with the restrictions to each of the individual open intervals where all the various restrictions are defined--i.e., on the intersection itself. A more relevant example building on the definitions just below is described in \S \ref{sec:cnfSat}.

A notion of locality in a space is furnished by a \emph{topology}, i.e., a family of \emph{open sets} defined so that any union or finite intersection of open sets is open. A \emph{basis} for a topology is a family of open sets whose unions generate the topology. One we have a topology, a notion of local data on a space is furnished by \emph{sections} defined over any open set. For example, as pointed out in \cite{rosiak2022sheaf}, a basis for a particular topology for Earth is the family of jurisdictions: then open sets are unions of jurisdictions, and sections are the laws over an open set. \footnote{The choice of topology is crucial for two main reasons: i) to avoid algorithmic scaling problems, and ii) to provide a strong set of constraints.
The point i) means that if too many collections of claims are too heavily connected, determination of whether they are consistent becomes infeasible. In essence, the topology (possibly mediated by an auxiliary sparsity-enforcing method) governs which sets of claims need to be tested for consistency, and which are likely to be unrelated, and so do not need to be tested.
The point ii) is also crucial, because it enables many inferential tools to be deployed. 
If the topology is too simple, the sheaves that can be constructed with it become degenerate.
Following the jurisdictional example a bit further, 
there is little information to be drawn about the interrelations between laws in different jurisdictions if there are only a few, mostly unrelated jurisdictions under consideration.
}

If these sections can be \emph{restricted} to smaller open sets coherently, then the resulting structure is called a \emph{presheaf}. If furthermore the sections are wholly determined by these restrictions in the sense that agreement of sections over open sets implies that they are restrictions of a common section over the union, then the resulting structure is called a \emph{sheaf}. 
\footnote{For more general and abstract variants of the sheaf construction, see \cite{maclane2012sheaves}, \S 10-11 of \cite{rosiak2022sheaf} (especially Example 294 therein), and \cite{srinivas1993sheaf}. These variants are intimately related to logic and even the notion of truth itself via the notion of a \emph{topos} \cite{goldblatt2014topoi,sep-truth-values}.}
Again, \S \ref{sec:cnfSat} gives an example.

Sheaves have recently been used to model dynamics of networked opinion expression and lying and to simulate consensus and control mechanisms \cite{hansen2021opinion} (see also \cite{ghrist2022network,riess2022diffusion}). Moreover, algorithms for determining maximal(ly consistent) sections \cite{praggastis2016maximal} and \emph{consistency radii} that measure the amount of agreement between local sections and accommodate noisy data \cite{robinson2020assignments} have recently been developed and applied to problems in sensor fusion \cite{robinson2017sheaves,robinson2019hunting,joslyn2020sheaf}, differential testing \cite{ambrose2020topological}, and comparison of local and global goodness of fit in statistics \cite{kvinge2021sheaves}. Finally, \emph{sheaf cohomology} \cite{ghrist2014elementary,rosiak2022sheaf} enumerates and describes obstructions to consistently ``gluing'' local data together, as schematically indicated in Figure \ref{fig:triangle}. 

This is exactly the sort of thing that we want in principle, though in practice some algorithmic and architectural nods to expediency may be required \emph{in silico}. One possibility is to learn a sheaf neural network \cite{hansen2020sheaf} using a sparse subset of data. 
\footnote{This embodies the idea that a reasonable and efficient approach to test consistency at larger scales is to test a few times \emph{versus} every time. This generalizes the idea of exploiting ``spatial'' sparsity of a graph of networked claims by learning a graph neural network \cite{ijcai2022p772,pmlr-v198-deac22a}.}  
On a separate note, it seems likely that consistency data need not be strictly real-valued, but might account for noise by using distributions, intervals, or some other more flexible data that can yield more robust results: for example, {\L}ukasiewicz logic naturally dovetails with the quantized ratings in \S \ref{sec:local}.\footnote{
Anticipating \S \ref{sec:cnfSat}, we note that constraint problems involving {\L}ukasiewicz logic are readily handled using standard tools \cite{soler2016bit,preto2023linking}. Moreover, a {\L}ukasiewicz semiring is conducive to deployment of message-passing algorithms \cite{aji2000generalized}. 
}
It may also be useful to consider binary consistency measures and use contrast-consistent search \cite{burns2022discovering,farquhar2023challenges} to answer questions along the lines of ``Is [claim A] logically consistent with [claim B]?'' Yet another possibility is to consider ternary consistency measures that allow for ambiguity without quantification.

Ultimately, identifying good tradeoffs between things like expressivity, robustness, computational efficiency, and utility will be critical to a useful realization of the ideas in this proposal. Techniques involving things like data representations, sparsity, and graph/sheaf neural networks will respectively inform these tradeoffs, as will the choice of a LLM, prompt engineering, and distillation (semantic and/or statistical) of LLM outputs.

\subsection{\label{sec:cnfSat}Boolean satisfiability and cellular sheaves}

\emph{Conjunctive normal form Boolean satisfiability} (CNF-SAT) is the archetype of a computational problem whose solutions are known to be easy to verify but are presumably hard to compute in general \cite{schoning2013satisfiability,knuth2015art}.
\footnote{I.e., CNF-SAT is the archetypal $\mathbf{NP}$-complete problem.}
Writing $\wedge$ and $\vee$ to respectively indicate logical $AND$ and $OR$, and an overbar \emph{\`a la} $\bar x$ to indicate logical $NOT$, a representative CNF-SAT formula is
$$(w \lor \bar x) \land (w \lor y) \land (x \lor \bar y) \land (x \lor y \lor \bar z).
$$
This has the solutions or \emph{satisfying assignments} $$(w,x,y,z) \in \{(TRUE,FALSE,FALSE,FALSE),(TRUE,TRUE,*,*)\},$$ where $*$ indicates a wildcard that can take the value $TRUE$ or $FALSE$. Each of these assignments of Boolean values to the variables involved causes the formula above to evaluate to $TRUE$.
\footnote{More generally, a propositional formula is CNF-SAT if it has the form $\wedge_j \left ( \vee_k \lambda_{jk} \right )$, where the \emph{literal} $\lambda_{jk}$ represents either an atomic proposition/variable or its negation (say, $x$ or $\bar x$). The terms $\vee_k \lambda_{jk}$ are \emph{clauses}. The case where $k$ always ranges over $\{1,\dots,K\}$ is called $K$-SAT. There is a fast algorithm for 2-SAT, but 3-SAT is $\mathbf{NP}$-complete.}

The relevance of CNF-SAT to our considerations stems from the fact that any problem in propositional logic can be efficiently transformed into CNF-SAT by introducing auxiliary variables. As a standard format for reasoning about propositional logic \emph{in silico}, CNF-SAT offers a natural conceptual framework for dealing with the consistency of linked propositions or claims.

\begin{figure}[h]
  \centering
\includegraphics[trim = 70mm 115mm 65mm 115mm, clip, keepaspectratio, width=.32\textwidth]{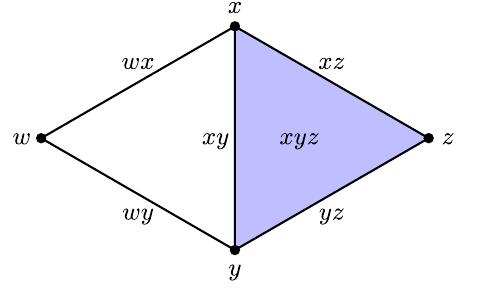}
\includegraphics[trim = 70mm 115mm 65mm 115mm, clip, keepaspectratio, width=.32\textwidth]{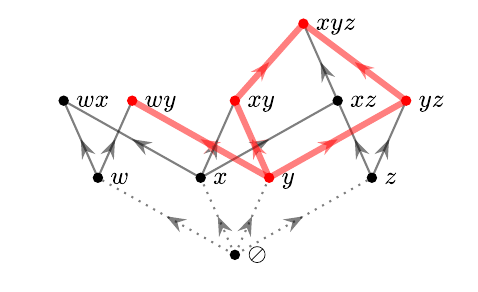}
\includegraphics[trim = 70mm 115mm 65mm 115mm, clip, keepaspectratio, width=.32\textwidth]{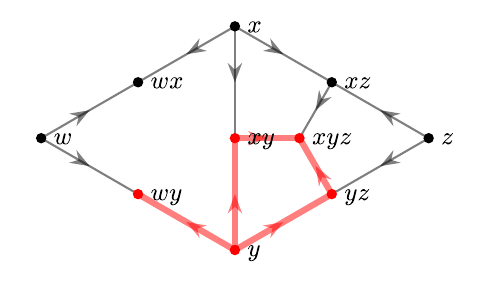}
  \caption{(Adapted from \cite{rosiak2022sheaf}; left) The CNF-SAT formula $(w \lor \bar x) \land (w \lor y) \land (x \lor \bar y) \land (x \lor y \lor \bar z)$ encodes an \emph{abstract simplicial complex} (ASC) whose constituent simplices correspond to (sub)clauses and are labeled according to the participating variables. (A dual ASC [not shown] switches the roles of clauses and variables but has the same gross structure: viz., \emph{simplicial homology} indicating a single hole \cite{dowker1952homology,ghrist2014elementary,huntsman2022topology}. It is a triangle formed on vertices $wx$, $wy$, and $xyz$ respectively connected by one-dimensional simplices labeled $w$, $y$, and $x$: some evident redundancy has been eliminated.) (Center) The inclusion of simplices defines a \emph{partial order} and a concomitant topology in which the open sets are unions of ``up-sets'' such as {\color{red}$\uparrow y$, which is highlighted in red}. (Right) Redrawing the partial order with the empty simplex $\varnothing$ omitted gives an ``attachment diagram'' that indicates how open sets indeed encode a sensible notion of locality in the ASC.}
  \Description{CNF-SAT formulas define cellular sheaves.}
  \label{fig:cnfSat}
\end{figure}

CNF-SAT is also the archetype of a \emph{constraint satisfaction problem}: every \emph{clause} in parentheses introduces a constraint that has to be individually satisfied in order for the larger formula to be satisfied. Meanwhile, adding (resp., deleting) clauses to a CNF-SAT formula generally reduces (resp., increases) the number of solutions. This was elegantly contextualized in \cite{srinivas1993final}:
\begin{quote}
the solutions to a constraint satisfaction problem form a sheaf: any consistent assignment must be assembled from consistent parts. Constraint satisfaction algorithms search for consistent assignments of values to variables.
\end{quote}
Indeed, enumerating the satisfying assignments to a CNF-SAT formula corresponds precisely to a particular sheaf cohomology computation, though the only concrete demonstration of this that we are aware of takes a highly technical detour into algebraic geometry \cite{bach1999sheaf}. 
\footnote{While this enumeration problem of \emph{model counting} is $\#\mathbf{P}$-hard, practical approximation algorithms exist \cite{chakraborty2021approximate}. Model counting is particularly relevant to establishing relative inconsistency measures \cite{besnard2020relative}. For example, given two CNF-SAT formulas $p$ and $q$, we might gauge their consistency with a relative measure such as the number of satisfying assignments (or models) of $p \wedge q$ divided by $2^V$, where $V$ is the number of variables in $p \wedge q$.}
Without taking this detour, Figure \ref{fig:cnfSat} still sketches how a CNF-SAT formula gives rise to a topology and a notion of restriction, and in turn to a sheaf whose global sections are precisely the satisfying assignments.

We briefly elaborate: because ASCs such as in the left panel of Figure \ref{fig:cnfSat} are special cases of so-called \emph{cell complexes}, the constructions involved yield a Boolean-valued \emph{cellular sheaf} \cite{curry2014sheaves,hansen2019toward,rosiak2022sheaf} along lines first suggested in \cite{srinivas1993sheaf}. If we take vertices and faces (of whatever dimension) to respectively correspond to variables and clauses, then sections over open sets (which are given by the so-called \emph{Alexandrov topology} indicated by the $\uparrow$ construction) are simply assignments of Boolean values to the local variables that satisfy the corresponding parts of the CNF-SAT instance. Typically, many if not most local sections will not extend to local sections over larger open sets, let alone extend to global sections.


As a practical matter, we will not and should not expect to find satisfying assignments of propositions or claims that are contradictory. However, it is possible to emphasize certain claims, e.g., any that are beyond reasonable dispute, and to replace the criterion of \emph{total} satisfiability with \emph{maximal} satisfiability, by analogy with \cite{praggastis2016maximal,robinson2020assignments}. The corresponding \emph{maximum satisfiability} (MAX-SAT) problem is to maximize the sum of weights associated with satisfied clauses \cite{bacchus2021maximum}. While this problem is $\mathbf{NP}$-hard, it admits an integer programming formulation whose linear programming relaxation is particularly efficient for small clauses \cite{kearney2020sheaf,vazirani2001approximation}, and perhaps of practical relevance for our considerations in its own right. 

A different perspective on the same ideas is that some assignments of propositions or claims are ``more satisfying'' than others. This is reflected in the structure of the \emph{consistency filtration} of \cite{robinson2020assignments} and quantified in the consistency radius of the sheaf. A small radius may simply be due to noise, but a large radius indicates an unresolvable inconsistency.


\section{\label{sec:deployment}From concept to deployment}

While any new technology faces a ``valley of death'' between research and adoption \cite{ellwood2022crossing}, a proposal such as ours would also inevitably face significant administrative and/or political challenges in many if not all of the application domains we have mentioned. Furthermore, the infrastructure required to deploy and scale a system such as we propose is expensive. In short, there are two main development gaps: between concept and demonstration, and between demonstration and deployment. Research and development organizations can address the former gap but are notoriously ill-equipped to address the latter gap.

Bearing in mind the role of (United States) government in technology development \cite{block2015state}, the realization of a system such as we propose is not easily imagined under (say) the Federal Judicial Center or National Institute for Justice, or for that matter under private organizations. The intersection of the remit of agencies that are well-equipped to bridge the first of the gaps above and of natural application domains is arguably centered on countering malign influence operations in the information-cognitive space \cite{waltzman2017weaponization,bay2019current,nemr2019weapons}.  

Meanwhile, a \emph{public-private partnership} (PPP) would be a reasonable vehicle for bridging the second gap. Existing research and best practices \cite{draxler2008new,huntsman2013private,state2019fam,state2024framework} argue for a broad coalition of like-minded partners, structured as a PPP, as the top approach to solve complex global challenges such as disinformation. Strategic long-term PPPs are required when a challenge is too great for government or private stakeholders to solve on their own. Desired outcomes are achieved by way of cross-sector integration of resources and expertise \cite{wef2005building,draxler2008new} as well as mobilization of competencies and commitments by public, business and civil society partners in order to achieve a strategic goal that cannot be achieved independently. Whereas legal partners share risk and benefit equally, partnerships for creation of public good---such as integrity and safety of information---are not always based on equality either of contribution, risk, benefit or losses \cite{gerrard2007sourcebook}: partners are equal in their status but the size of their contribution, risk and benefit typically varies. 

In the present context, including stakeholders that are targets of cognitive warfare is crucial to enable a better understanding of the operating environment \cite{altenburg2005private}, diverse perspectives, and building institutional capacity \cite{draxler2008new}. This suggests a role for international partners in a PPP \cite{wef2024global} and the prospect of bridging a third gap between initial defensive deployments and subsequent deployments focused on improving governance across liberal democracies.

\section{\label{sec:conclusion}Conclusion}

There are many recent and surely also ongoing investigations of how to get LLMs to answer questions about facts \cite{tian2023fine,wang2023survey}, as well as how to cope with outright lying by a LLM itself \cite{azaria2023internal,liu2023cognitive,marks2023geometry,park2023ai}. However, it is hard to imagine how LLMs or any other generative artificial intelligence tools could ever be trustworthy arbiters of truth, because they fundamentally rely on sampling from a probability distribution that itself depends on  ``received wisdom.'' There may be technical mitigations, e.g., performing what amounts to cross-validation \cite{dhuliawala2023chain}. However, trust is ultimately social and not necessarily even moored to facts or truth, as common phenomena such as confirmation bias, cognitive dissonance, doublethink, hypocrisy, and receptivity to bullshit \cite{pennycook2020falls} amply demonstrate.

We avoid this trap by focusing on logical consistency instead of truth \emph{per se} \cite{mundler2023self,li2023contradoc}, and also allowing for the possibility of errors or noise in claims. That said, we could summarize the present proposal as arguing that it is now possible to begin algorithmically instantiating a coherence theory of truth \cite{sep-truth-coherence}. This does not preclude, and could be well complemented by, taking conclusively established facts as partial ``boundary conditions'' to anchor the coherence theory in a correspondence theory of truth \cite{sep-truth-correspondence} to the extent practicable, i.e., to try reduce a space of global or maximal sections to those that are also consistent with observed reality. The technical apparatus for such a step is established in \emph{retrieval-augmented generation} \cite{lewis2020retrieval,gao2023retrieval}, but given that many established facts conflict with other legally protected and widely held beliefs, liberal democracies would encounter obstacles here even with flawless technology. The roles of culture and language in epistemic disconnects (as indicated by experiments in cultural transmission along the lines of games like Chinese whispers or telephone \cite{mesoudi2008multiple} and the analogue of round-trip translation) also require delicacy in handling that technology is unlikely to provide on its own. Indeed, \cite{perez2005institutionalization} points out that in a pluralistic society, consistency in law may come at the cost of fairness.

Ultimately, many citizens will always disagree with others about basic facts, let alone matters of opinion. We can only reasonably hope that citizens get reliable tools to identify where their basic disagreements actually lie.

\begin{acks}
Thanks to George Cybenko, James Fairbanks, Thomas Gebhart, Neil Gerr, Letitia Li, Daniel Rosiak, and Wlodek Zadrozny for helpful conversations and commentary.

\end{acks}

%
\bibliographystyle{ACM-Reference-Format}
\bibliography{biblioDGOV}

\appendix

\section{\label{sec:prompt}Initialization prompt used for results of \S \ref{sec:local}}

\begin{quote}
\texttt{Imagine you are a perfectly objective arbitrator with impeccable judgment and integrity. In response to a prompt of the form ``evalConsistency: '' followed by two claims in square brackets that are separated by a space, please do two things. First, explain the logical relationship between the two claims, assuming that the first claim is true, whether or not it actually is. I want you to ignore the truth, falsity or basis in fact of either claim. Second, use your explanation to numerically rate the relative consistency of the two claims. Do not pay attention to or comment on the truth or basis in fact of either claim independent of the other. Your rating of relative consistency should be on a scale from 0 to 10, with a value of 0 for a pair of claims that are not at all consistent and a value of 10 for a pair of claims that are totally consistent. I cannot emphasize enough that for your rating, I want you to ignore the truth or basis in fact of either claim, since anything that is not consistent with reality cannot be true. To be clear, a pair of unrelated claims should be rated a 10 and a pair of false but consistent claims should also be rated a 10. Meanwhile, a pair of claims of which one is true and the other is false, should be rated a 0. Your response must end with the numerical rating.}

\

\texttt{For example, the prompt}

\

\texttt{``evalConsistency: [The earth is flat] [The sky is red]''}

\

\texttt{should produce a response like}

\
 
\texttt{``The shape of the earth and color of the sky are unrelated, so the consistency rating of these claims is 10.''}

\

\texttt{As another example, the prompt}

\

\texttt{``evalConsistency: [Purple people are evil] [Purple people are good]''}

\

\texttt{should produce a response like}

\

\texttt{``If either claim is true, then the other is false, so the consistency rating of these claims is 0.''}

\

\texttt{Your response must end with the numerical rating.}
\end{quote}

See also \url{https://github.com/SteveHuntsman/ProspectsForInconsistencyDetection}.

\end{document}